# Terrestrial effects of moderately nearby supernovae

Adrian L. Melott[1] and Brian C. Thomas[2]

*Abstract-*

Recent data indicate one or more moderately nearby supernovae in the early Pleistocene, with additional events likely in the Miocene. This has motivated more detailed computations, using new information about the nature of supernovae and the distances of these events to describe in more detail the sorts of effects that are indicated at the Earth. This short communication/review is designed to describe some of these effects so that they may possibly be related to changes in the biota around these times.

1. Adrian L. Melott (melott@ku.edu) Department of Physics and Astronomy, University of Kansas, Lawrence, Kansas 66045 USA.
2. Brian C. Thomas (brian.thomas@washburn.edu) Department of Physics and Astronomy, Washburn University, 1700 SW College Ave., Topeka, Kansas 66621 USA

**Running head: Supernova effects**

## Introduction

Recently, there has been a rush of new evidence concerning supernova events moderately near to the Earth. Supernovae have been considered as a possible cause of terrestrial extinctions and lesser events for a long time, at least since Schindewolf (1954). Later work was done by Ruderman (1974); Whitten *et al.* (1976); Gehrels *et al.* (2003); Horvath and Galante (2012); Thomas et al. (2008); Melott et al. (2015, 2017). Due to recent detections, there have been new computations which shed light on the effects of events at moderate distances, such as 150 to 300 light years. Such events should occur much more frequently than the mass extinction level events at about 30 light years or less which are expected only at intervals of several hundred My. The purpose of this brief review for the paleontological community is to summarize the nature of the expected effects, some of which are different than considered before, and greater than previously thought for the moderate distance events. The effects are somewhat disparate in their physical and chemical mechanisms as well as the type of effect on the biota; what they have in common is that they are likely to be important from supernovae at moderate distances, in particular the event(s) in the late Miocene and the early Pleistocene.

## Supernovae

It is not necessary to discuss in any detail the nature of supernovae, other than to mention that they occur in two main types, imaginatively called Type I and Type II. Type I is primarily a result of a small object, such as a white dwarf star, gravitationally robbing gas from a companion until it becomes too massive to support itself against gravity, and collapses. Type II usually results at the end of the lifetime of a very massive star (much larger than the Sun). As this type exhausts its fuel, it will swell up and then suddenly collapse. The rebound results in a blast of light, radiation, and a shell of ejected material. The typical released kinetic energy for the two types is about the same.

Including both types, there are about three supernovae per century in our galaxy. Many of these cannot be seen due to dense intervening gas. There have been a number noted within historical times (Stephenson 2016), and in the cases of most of them we can still see a remnant, in the form of an expanding shell of hot gas. Most of these have been so bright that they were seen in the daytime, despite being as much as thousands of light years away. So it is easy to see that there may be substantial effects from events ten or more times closer. Our purpose here is to describe those effects, so that they may possibly be related to effects seen in the fossil record. Recently, researchers have found concrete evidence for events a few Myr old (Melott 2016). Supernova remnants can only be seen in visible light for up to 100 kyr maximum. Recent interest is based on new isotopic evidence, a very broad field (pioneered by Fields and collaborators) which cannot be covered here. For a more complete and timely discussion of possible isotopic evidence, see Fry *et al*. (2016); see also Ludwig *et al*. 2016. At the present time there exists strong evidence for at least two events: late Miocene (7-8 Ma) and most significantly near the Pliocene-Pleistocene boundary (~2.5 Ma; Melott 2016). This evidence is based on multiple significant detections of $^{60}$Fe. However, we will focus only on generalized features expected from such events, using these instances as examples. In particular, the more recent event is estimated to have taken place at a distance of about 150 light years (Knie *et al*. 2004; Fry *et al*. 2016; Mamajek 2016).

**Computing terrestrial effects of supernovae**

In our recent papers (Thomas *et al*. 2016a; Melott *et al*. 2017) we discussed the likely terrestrial effects of supernovae at distances of 150 to 300 light years. Such events are likely (based on the average rate in the galaxy) to come along every few Myr, whereas the extinction-level events at 30 light years or closer are only likely at intervals of a few hundred Myr (Ellis *et al*. 1996). It is important to note that a steady rate is not expected. Most of the supernovae we are likely to encounter are from very massive stars, perhaps 10 solar masses. Most such stars are born in large gas clouds with a considerable number formed at the same age. The most massive stars have the shortest lifetimes (due to their rapid use of fuel), so that large stars have short lives, perhaps only a few Myr. They tend not to move very far from their siblings before exploding, so the Earth is likely to go for a substantial amount of time without supernova effects, then be affected by many of them at close to the same time.

Our work has been based on new information about supernovae, some of it resulting from new technology which allows them to be observed very soon after the initial explosion. There are a number of possible effects. One is being blasted by (potentially radioactive) dust. This is likely to be a hazard only in the case of the infrequent very close events, but is the basis for recent detections at low levels which have confirmed one set of events (Ellis *et al*. 1996; Fry *et al*. 2016).

The type IIP (a subtype of type II supernovae), which dominate the likely sources, are very weak in high-energy electromagnetic radiation, variously called gamma rays, X-rays, and ultraviolet light. These direct burst effects are insignificant at the distances we are considering (Melott *et al*. 2017). However, visible light effects may be somewhat significant. An event at 150 light years would be about as bright as the full moon. Since it is a point source, it would be visible in the daylight. If it were on the night side of the Earth, it would light up the sky. Such objects begin as very blue, and gradually fade in brightness and color toward the red. There is a large literature, summarized in Melott *et al*. (2017) about the deleterious effects of nighttime blue light on living organisms. There are hormonal and behavioral effects. Such light can have insomnia-inducing effects on humans and even enhance cancer rates. However, this phase is likely to last only about two weeks, and so would be undetectable in the fossil record.

The more important group of effects, and those of greater intensity than thought in the past, are based on cosmic rays. Cosmic rays are the nuclei of atoms, moving at very high speeds, close to the speed of light. They constantly impinge on the Earth. However, supernovae are thought to be the primary source of cosmic rays within our Galaxy, so a nearby event can understandably boost the flux of these particles a great deal. Energy and mass are interchangeable, and it is customary to characterize the speed of such relativistic particles by their total energy.

We have undertaken a series of computations, the most relevant here published as Melott *et al*. (2017). In this, we have used the spectrum of cosmic rays inferred for a typical type IIP event. We have computationally propagated the cosmic rays through several alternate possible models for the magnetic field of the galaxy between the supernovae and the Earth. Then, additional computations follow the showers of radiation down through the atmosphere, to find the flux of various kinds of radiation on the ground. Other computations have followed the ionization of the atmosphere induced by the radiation as it passes through the atmosphere, and then atmospheric chemistry computations follow the modifications of the composition of the atmosphere. This represents four different substantial complex computations and it is beyond the scope of this summary to describe them in detail, as is initiated in Melott *et al*. (2017) and references therein. We summarize and describe the effects of primary importance for biology.

The kinetic energy of cosmic rays produced by supernovae are up to a million times their mass, and the flux of such particles from a supernova at 150 light years can be increased by a factor of 200 over the customary background (Melott *et al*. 2017). The paths of cosmic rays are deflected by magnetic fields in the Galaxy, so they do not travel in straight lines to the Earth, but rather diffuse. What began at the supernova as a rather quick burst will be spread out over time—taking perhaps 10,000 yr for the levels to return to normal. The recently confirmed event appears to have taken place inside the Local Bubble—so the cosmic rays will essentially be trapped by the walls of this bubble and reflected back by its intense magnetic field, causing the irradiation of the Earth to persist for a long time, perhaps hundreds of thousands of years. Therefore we will emphasize steady-state effects.

The intense flux of cosmic rays at the Earth will have several effects. They strike the atmosphere, producing "showers", as the high energy particles produce other particles, all heading downward toward the Earth. Normally, we get only a small amount of radiation from this, as the atmosphere provides substantial shielding.

**Ozone depletion and UVB increase**

For a long time the primary effect of radiation that has been considered is ozone depletion (Reid *et al*. 1978; Scalo & Wheeler 2001). Most cosmic rays deposit their excess energy in or below the stratosphere, by breaking up molecules and knocking electrons out of atoms. Normally, as is well known, nitrogen is all tied up in the very stable $N_2$ molecule, and not available for life. Available nitrogen, like nitrates, thus makes powerful fertilizer. The cosmic rays break up the $N_2$ molecule, and the isolated N become available to react—usually with the first thing they encounter. Since much of the atmosphere is oxygen, a lot of oxides of nitrogen are produced. Photochemical smog formerly contained large amounts of nitrogen dioxide produced in automobile engines. This product, however, is produced primarily in the stratosphere by cosmic rays. This process is described in detail in Thomas *et al*. (2005).

The nitrogen dioxide acts as a catalyst and converts ozone ($O_3$) back to ordinary oxygen ($O_2$) (Thomas *et al*. 2005). It is recycled in this process. It takes 5-10 years for the $NO_x$, as they are called, to be removed from the atmosphere, primarily by being rained out after a burst event. As this happens, it will provide an enhanced fertilizer for the nitrate-starved biota (Thomas & Honeyman 2008; Neuenswander & Melott 2015). However, with a relatively steady source, such as the flux of cosmic rays from a nearby supernova, they are constantly produced and reach a steady-state. The ozone depletion does too, and in our recent example computations showed a nearly steady mean 25% depletion in stratospheric ozone from the 150 light year supernova. This can be compared with a 3-5% maximum global average ozone loss in the late 20$^{th}$ century from refrigerants. Such depletion is latitude-dependent, varying in a complex way but generally worse near the poles than the equator (Neale & Thomas 2016).

The reason that stratospheric ozone depletion is serious has to do with ultraviolet radiation. The ozone layer in the stratosphere absorbs in a band called UVB which is responsible for skin cancer, cataracts, and is dangerous for unicellular organisms. The previously mentioned 25% depletion would produce about a 50% average increase in UVB at the surface (Figure 1). Unicellular organisms are effectively transparent to it, and their DNA is directly irradiated by it. UVB is energetic enough to break chemical bonds in biological molecules, so it can cause a great deal of damage. On the other hand, even simple organisms have molecular mechanisms that repair damage to DNA, so the overall effect is complicated. Experimental work with two important ocean phytoplankton showed a fairly small effect on primary productivity (Neale & Thomas 2016), but those experiments were short-term and do not address long-term survival. Analysis of UV damage to other types phytoplankton show a wide range in the severity of impact (Thomas *et al*. 2015).

Severe events of this type have been proposed as a candidate for initiating major mass extinctions, through severely damaging the base of the food chain in the oceans (Melott & Thomas 2009). Such events lack the usual geochemical signatures associated with other extinction causes, such as sulfur compounds, mercury, iridium, oxygen isotope variations, etc. However, the coincidence of supernova indicators such as $^{60}$Fe with an extinction event about 2.6 Ma is a powerful clue that an association may be present. Spores tend to develop a UV-resistant coating in response to elevated levels, and it is durable over geological timescales. Sporopollenin is a powerful proxy for UVB levels (Fraser *et al*. 2014), however it is not unique to astrophysical causes (Thomas *et al*. 2016b). The presence of a sporopollenin increase with no associated chemical cause for elevated UVB such as volcanism may be taken as indicative, but not proof, of a probable astrophysical cause. UVB direct effects would be confined to the surface and the upper few meters of water depth.

**Effects of increased UVB**

Increased UVB radiation is damaging to all types of organisms. However, the magnitude of impact varies widely depending on particular species and effect considered. Thomas *et al*. (2015) performed detailed radiative transfer modeling combined with "weighting functions" that quantify biological impact of a change in UVB irradiance; we can use those results to make some estimates for the supernova case. More detailed analysis of photobiological effects in this case can be found in Thomas (2017). In the most realistic case presented in Melott *et al*. (2017), $O_3$ column density is reduced by 15-40% depending on latitude. Under these conditions, UVB irradiance is increased by about 50% on average (locally higher or lower), increasing skin damage (sunburn, known as "erythema") and skin cancer risk by a similar factor. Figure 1 shows the increase in UVB irradiance following onset of cosmic ray-induced ozone depletion; time 0 indicates an arbitrary reference time after steady-state atmospheric conditions are achieved.

Marine phytoplankton productivity would be reduced, but the magnitude varies widely depending on species. For species commonly found in today's Antarctic ocean, reductions of 5-20% can be expected. Figure 2 shows the increase in UV-induced reduction in Antarctic phytoplankton productivity, using a biological weighting function from Boucher *et al*. (1994); as in Figure 1, time 0 indicates an arbitrary reference time after steady-state atmospheric conditions are achieved. For mid-latitude species the reductions are smaller (Neale & Thomas 2016). Damage to land plants could be increased by 10-80%, again depending on species. By comparing to existing field observations and experimental studies, Thomas (2017) concluded that UV damage is unlikely to cause widespread extinction but is likely to contribute to changes in ecosystem balances, with significant impact on some species with little negative (and possibly even positive) impact on others.

An important point is that the UVB enhancement persists for at least thousands of years, due to the cosmic ray confinement in the Local Bubble, much longer than any such change previously considered. The biological results that are known (including those discussed above) are based on relatively short-term studies ranging from hours to years. It is unknown how, for instance, phytoplankton survivability is impacted by long-term exposure; this is an area requiring further research.

**Other ionizing radiation at the surface**

One of the effects recently found to be more substantial than previously thought for "moderate" distance supernovae is radiation on the ground. One of the elementary particles produced in abundance in air showers is muons. A muon is similar to an electron, but with about 207 times greater mass. The greatest source of muons under normal conditions is cosmic rays, and they contribute about a sixth of the radiation normally encountered on the ground. They are extremely penetrating because they do not interact strongly with ordinary matter (Marinho *et al*. 2014). However, they are present in such great abundance that they have a substantial effect; due to their penetration they can have effects on biota as deep as a kilometer in water or several hundred meters underground. Most forms of radiation to which we are exposed cannot do this; internal doses can come, for example, from inhaling radon gas. Neutrons are also produced in abundance, but they mostly do not reach the ground. They are important at "airline" altitudes, which are essentially irrelevant for paleontology.

Recent computations modeled for the first time the specific flux of muons on the ground expected from a moderately nearby supernova (Melott *et al*. 2017). The ground level radiation dose from muons could easily rise by more than a factor of 150 from about 0.2 mSv per year to about 30 mSv per year (Melott *et al*. 2017) taking of order 10,000 yr to return to normal from an event without assuming cosmic ray confinement in the Local Bubble. If there is confinement, this dose could last for 100,000 yr. Although this dose sounds large, it is still only enough to increase the cancer risk for long-lived organisms by of order 5% per year of exposure. Mutation rates should be similarly affected.

The signature of muon exposure, different from all other radiation loads, would be a bigger effect on large organisms. As they are effectively nearly transparent to muons, the dose would be proportional to their mass, not their surface area; so the dose per unit tissue mass would be constant for all organisms. Normally, most radiation effects do not penetrate very deeply (except for radon inhalation). Therefore the dose would be proportional to surface area, and the dose per unit mass would be smaller for the largest organisms. So, proportionately, the biggest increase in radiation effects from muons should be seen in megafauna. Also, unlike UV effects, deep penetration into water is important. This makes a newly-identified late Pliocene marine megafaunal extinction (Pimiento *et al*. 2017) of considerable interest.

**Atmospheric ionization effects**

The mechanism of the aforementioned ozone depletion is due to ionization of the atmosphere. The large energy of the cosmic rays will knock electrons out of atoms as well as break up molecules. This has been known for decades (Reid *et al*. 1978), and discussions of terrestrial supernova effects have emphasized it. There is, however, an important difference in the cosmic rays expected to arrive from supernovae: their energy.

At the energies where the bulk of normal cosmic rays arrive at the Earth, the supernova is only expected to enhance the number of them by a factor of ten, at most. But supernovae are thought to generate abundant cosmic rays with energies a million times larger—and those will be enhanced in number by factors of order a few hundred. While the bulk of other kinds of space weather and even gamma-ray bursts ionize the upper stratosphere (30-40 km being typical altitudes for peak effects) these high energy cosmic rays penetrate much deeper. The peak ionization rate is at 10 km, deep in the troposphere, where weather takes place. Even down at the surface, ionization rates are up by close to a factor of 100. This is a new kind of effect, and we can only begin to describe what is likely to happen.

Cosmic rays have been proposed to play a central role in the initiation of lightning (e.g. Gurevich & Karashtin 2013). A voltage difference builds up in the atmosphere, but it is difficult for a current to get started until there are some free electrons. A cosmic ray shower passing through will provide that; a current begins to flow. That current knocks free more electrons, and the result can build up to a lightning strike. With a huge increase in the number of cosmic rays traversing the lower atmosphere to the ground, we can expect a similarly large increase in the amount of lightning, especially cloud-to-ground lightning.

Major climate modeling will be needed to disentangle the effects of such an increase, but it seems clear that greatly increased lightning will increase the rate of wildfires, with potential to transform terrestrial vegetation.

## Summary

It has been known for some time that moderately nearby supernovae may have substantial effects on the Earth. Events at ~150 light years will happen on average every few Myr, but will tend to happen in groups, with long periods between with no events. The effects of cosmic rays from such events appears to be greater than estimated previously. Ozone depletion and the increase of hazardous UVB continues to be important, but new effects come to the fore. Muon irradiation on the ground and hundreds of meters down into the ocean will increase cancer and mutation rates, the differences being most notable in terrestrial megafauna and benthic organisms. Typically larger organisms live long enough to develop cancer; in microorganisms the primary effects would be associated with mutation rates. Atmospheric ionization in the troposphere will greatly increase lightning rates, with a concomitant increase in the rate of wildfires.

## Acknowledgments

Helpful comments were provided by Bruce Lieberman, Richard Bambach and two anonymous referees. Research support was provided by NASA under research grant NNX14AK22G. There are no conflicts of interest.

Figure 1

Percent difference in UVB irradiance, as a function of latitude and time for the SN case (versus control). Time 0 indicates an arbitrary reference time after steady state is achieved.

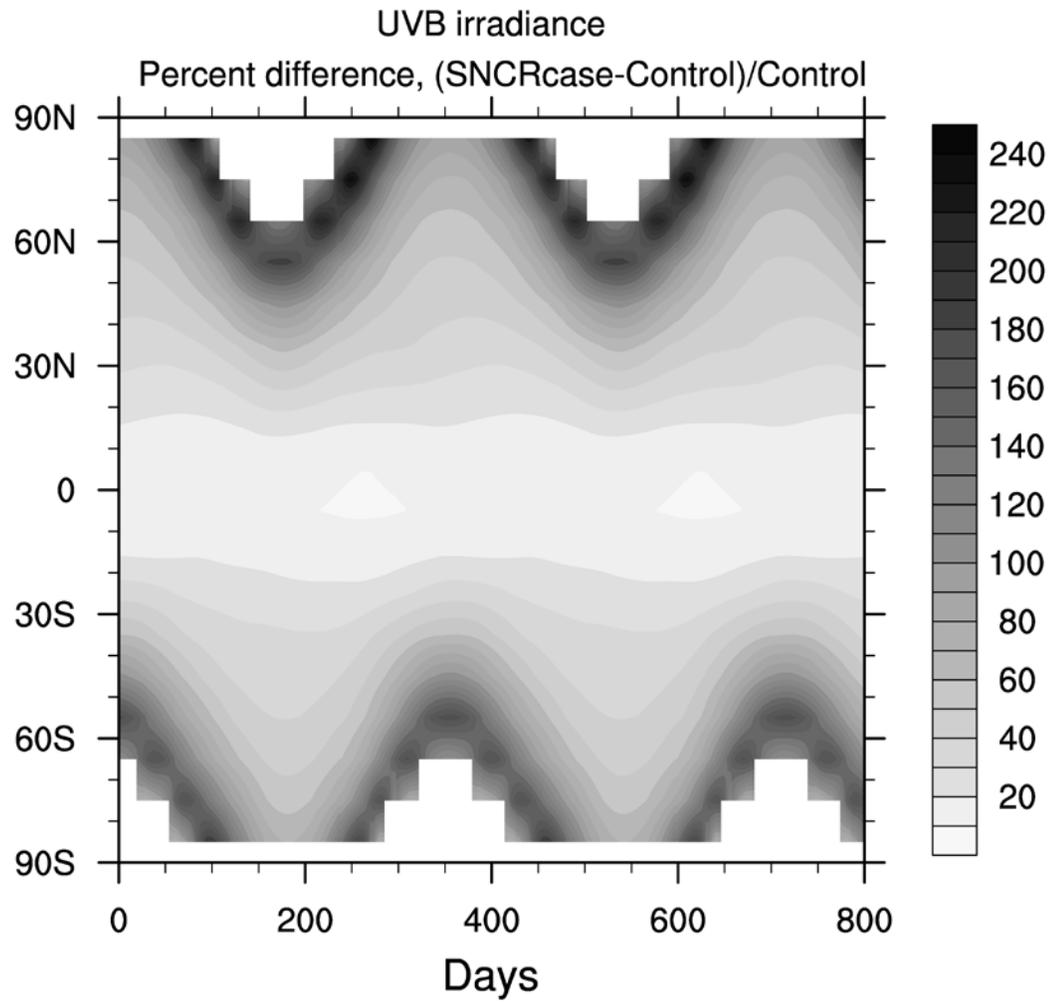

Figure 2

Percent difference in inhibition of phytoplankton primary productivity by UV, using weighting function from Boucher *et al*. (1994), as a function of latitude and time for the SN case (versus control). Time 0 indicates an arbitrary reference time after steady state is achieved.

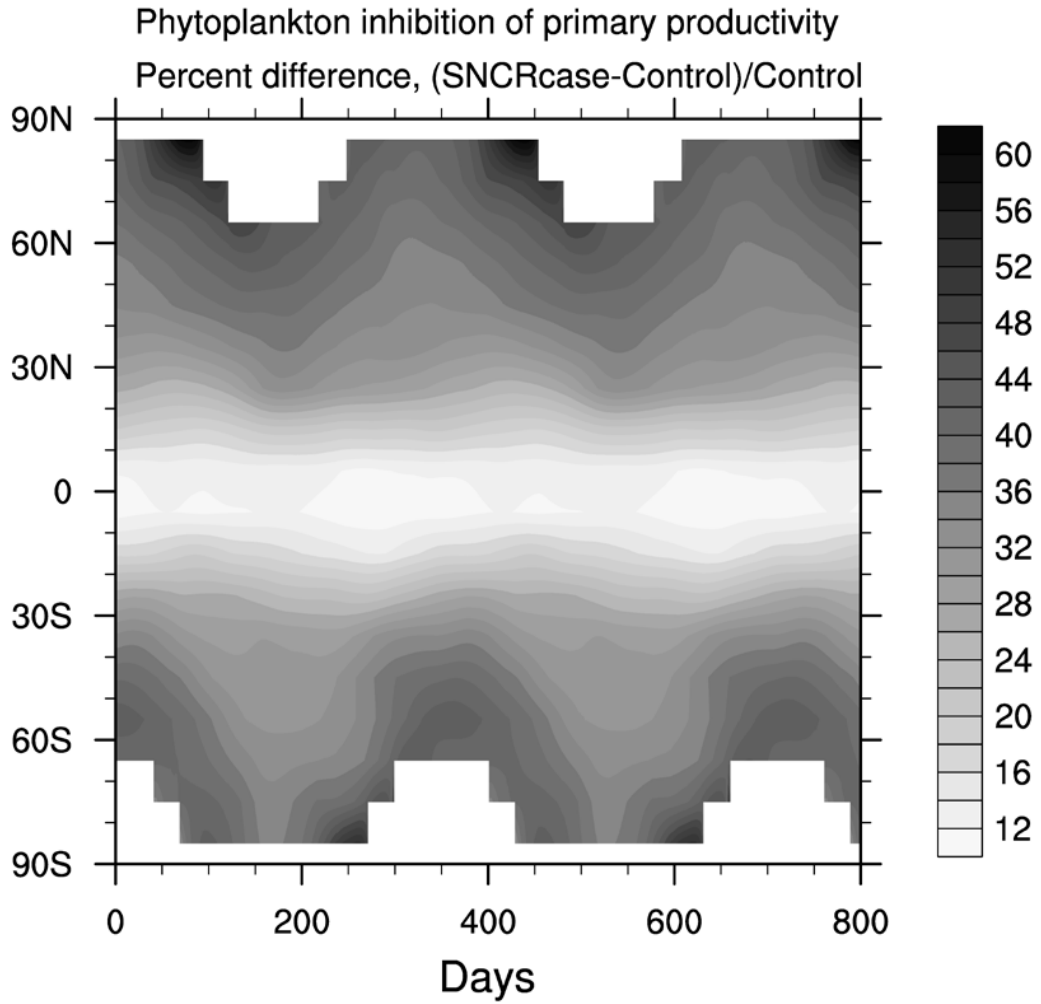